\documentclass[10pt]{npqcd}
\begin{document}
%Do not modify the line below:
\renewcommand\nextpg{\pageref{pgs1}}\renewcommand\titleA{
Quark Gluon Plasma Bags as Reggeons 
}\renewcommand\authorA{
%Authors with the references to e-mails:
K. A. Bugaev 
}\renewcommand\email{
%E-mails:
 e-mail:\space \eml{}{bugaev@th.physik.uni-frankfurt.de}\\[1mm]
}\renewcommand\titleH{
%Title for headers:
Quark Gluon Plasma Bags as Reggeons 
}\renewcommand\authorH{
%Authors for headers:
Bugaev K. A.
}\renewcommand\titleC{\titleA}\renewcommand\authorC{\authorH}\renewcommand\institution{
%Author institution:
Bogolyubov Institute for Theoretical Physics, Kiev, Ukraine
}\renewcommand\abstractE{
Within an exactly solvable model I discuss the  influence  of the  medium dependent finite width of quark gluon plasma (QGP) bags on their equation of state. 
It is shown  that the  large width of the QGP bags  not only explains 
the observed deficit in the number of  hadronic resonances, but also clarifies the reason   why 
the heavy QGP bags   cannot be directly observed  as metastable  states in a hadronic phase. 
I show how the model allows one  to estimate the minimal 
value  of the width of QGP bags  being heavier than 2.5  GeV 
 from a variety of the lattice QCD data and to  get  
the minimal  resonance width at zero temperature  of  about 600 MeV.
The Regge trajectories of  large and heavy  QGP bags  are established both in a vacuum and in a strongly interacting medium. 
It is  shown  that    at  high  temperatures the average mass and width of the QGP bags  behave in accordance with   the upper bound of the Regge trajectory asymptotics (the linear asymptotics), whereas at low temperatures (below a half of the Hagedorn temperature $T_H$ \cite{Bugaev:Ref0}) they obey  the lower bound  of   the Regge trajectory asymptotics (the square root one). 
Thus, for  temperatures below $T_H/2$ the spin of the QGP bags is restricted from above, whereas 
for  temperatures above  $T_H/2$  these  bags demonstrate the typical Regge behavior consistent with the string  models.  
}

\begin{article}

\section{Introduction}
Recently several ground shaking results in  statistical mechanics of  strongly interacting matter are obtained \cite{Bugaev:Ref1,Bugaev:Ref2,Bugaev:Ref3,Bugaev:Ref4, Bugaev:Ref5}.  All of them are  derived within the exactly solvable models.
In \cite{Bugaev:Ref1,Bugaev:Ref2,Bugaev:Ref3,Bugaev:Ref4,Bugaev:Ref5}
 the surface tension of large QGP bags was  included into the statistical  description of the QGP  equation of state (EOS).
Such an  inclusion not only allowed one to formulate  the  analytically solvable statistical models for the QCD tricritical 
\cite{Bugaev:Ref1} and critical \cite{Bugaev:Ref5} endpoint,  but also  to  develop 
 the finite width model of QGP bags 
\cite{Bugaev:Ref2,Bugaev:Ref3,  Bugaev:Ref4} and to  bring  up the statistical description of  the QGP EOS  to a 
qualitatively new level of  realism by establishing the Regge trajectories of heavy/large QGP bags both in a vacuum and in a medium 
\cite{Bugaev:Ref3,  Bugaev:Ref4} using the lattice QCD data.   

Here I would like to summarize the main results obtained in \cite{Bugaev:Ref1,Bugaev:Ref2,Bugaev:Ref3,Bugaev:Ref4, Bugaev:Ref5} and discuss the theoretical and experimental  perspectives to study the properties of  the 
strongly interacting matter EOS.  

\section{From the MIT Bag Model EOS to Finite Width Model}

From its  birth and  up to now  the MIT bag model \cite{Bugaev:Ref6} remains one of  the most popular 
model EOS to describe the confinement phenomenon  of the color degrees of freedom. 
It gives a very simple picture of a color confinement by considering the  freely moving massless quarks and gluons inside the  bubble  with negative  pressure, $-B_{bag}$, (for vanishing baryonic charge). Then the total pressure of a bag is 
\begin{equation}\label{Bugaev:1}
p_{bag}~ \equiv ~ \sigma \, T^4 - B_{bag} \,, \quad \varepsilon_{bag}~ \equiv  ~T \frac{d \, p_{bag}}{d~T} -  p_{bag}~ =~  3 \, \sigma \, T^4 +  B_{bag} \,,
\end{equation}
where $3 \, \sigma $ is the usual Stefan-Boltzmann constant which can be  trivially  found for a given number of elementary degrees of freedom.  The non-trivial  feature of the bag model EOS  is that the vacuum pressure, $- B_{bag}$, generates  positive and large  contribution to the bag energy density  $\varepsilon_{bag}$.

In addition to  a  simplicity  of  the EOS (\ref{Bugaev:1})  its other  attractive features are based on the fact that  the bag model also  leads \cite{Bugaev:Ref7}  to the Hagedorn mass spectrum   of bags  
which  was the  first signal  about the new physics above the Hagedorn temperature $T_H$.  
The easiest way to understand a special role of the Hagedorn mass spectrum, i.e. an exponentially increasing with mass $m$ density of hadronic states  $\rho_H (m) \sim exp\left( \frac{m}{T_H}  \right)$   for $m \gg T_H$,  is to consider the microcanonical ensemble of the test particles being in a thermal contact with the resonance of the exponential spectrum \cite{Bugaev:Ref8}.  Then one can easily show that  
the exponential spectrum  behaves as the   perfect thermostat and   perfect particle reservoir, i.e.  it imparts its temperature $T_H$ to  particles which are in thermal
contact with it  and forces  them to be in chemical equilibrium \cite{Bugaev:Ref8}. 
Therefore there is simply  no reason to study  such a system in the canonical ensemble or in grand canonical ensemble, i.e. to bring it into the contact with an external thermostat that has other temperature than $T_H$, since two thermostats of different temperatures cannot be in equilibrium. 
Perhaps these properties  can explain the fast  chemical equilibration of hadrons in an expanding fireball  \cite{Bugaev:Ref9}.

The first successful statistical model explaining the deconfinement  phase transition (PT) from hadronic matter to QGP,   gas of bags model (GBM)  \cite{Bugaev:Ref7},  is based on the MIT bag model and  it  interprets such a PT as a formation of   the infinitely large  bag. 
Further development in this direction led to many interesting findings
 \cite{Bugaev:Ref10, Bugaev:Ref11, Bugaev:Ref12}, including an exact analytical solution for finite systems with  a PT   \cite{Bugaev:Ref10},  but the most promising results were  obtained only recently. 
The most hopeful  of them are  an inclusion of 
the quark gluon bags surface tension into statistical description  \cite{Bugaev:Ref1, Bugaev:Ref5}  
and   taking into account  the finite width of large/heavy QGP bags  \cite{Bugaev:Ref2, Bugaev:Ref3, Bugaev:Ref4}. Thus, the  surface tension of QGP bags introduced in the quark gluon bag with surface tension model (QGBSTM)   allows one to  simultaneously 
describe the 1-st and 2-nd order deconfinement PT  with the cross-over and  generate 
the tricritical  (QGBSTM1) \cite{Bugaev:Ref1}  or critical endpoint   (QGBSTM2)  \cite{Bugaev:Ref5}  at the vanishing value of the  surface 
tension coefficient as required for the PT  of  liquid-vapor type  \cite{Bugaev:Ref13,Bugaev:Ref14}.
The finite width model (FWM)   \cite{Bugaev:Ref2, Bugaev:Ref3, Bugaev:Ref4}  naturally 
resolved {\it  two conceptual problems} which, so far,  were ignored  by other   statistical approaches 
for almost three decades:
on  one hand,  the volume dependent   width  of large QGP  bags  easily 
explains 
a huge deficit in  the  number of observed  hadronic resonances \cite{Bugaev:Ref15}  with masses above 2.5 GeV predicted by the  Hagedorn model \cite{Bugaev:Ref0}  and used, so far,  by the  GBM and all  its followers; and, on the other hand,  the  FWM  shows that   there is an inherent property of the strongly  interacting matter  EOS, the {\it subthreshold suppression},  which prevents the appearance of  large  QGP bags and strangelets   inside of the  hadronic phase even as metastable states in finite systems which are studied in nuclear  laboratories.  The latter effect explains the negative results of   searches for strangelets and for not too heavy QGP bags, say with the mass of $10-15$ GeV,  in various physical  processes.

The most convenient way to study the phase structure  of  any statistical  model similar to the GBM or QGBSTM  is to use the isobaric partition \cite{Bugaev:Ref1, Bugaev:Ref5} and find its rightmost singularities. Hence,  I  assume that after the Laplace transform  the  FWM grand canonical  partition  $Z(V,T)$ generates the following 
isobaric partition:
\begin{equation}\label{Bugaev:Eq2}
\hat{Z}(s,T) \equiv \int\limits_0^{\infty}dV\exp(-sV)~Z(V,T) =\frac{1}{ [ s - F(s, T) ] } \,,
\end{equation}
\noindent
where the function $F(s, T)$ contains the discrete $F_H$ and continuous $F_Q$ mass-volume spectrum  of the bags 
\begin{equation}
F(s,T) \equiv F_H(s,T)+F_Q(s,T) = \sum_{j=1}^n g_j e^{-v_js} \phi(T,m_j) = 
+\int\limits_{V_0}^{\infty}dv\hspace*{-0.1cm}\int\limits_{M_0}^{\infty}
 \hspace*{-0.1cm}dm~\rho(m,v)\exp(-sv) \, \phi(T,m)~.
\label{Bugaev:Eq3}
\end{equation}
\noindent
The   density
of  bags of mass $m_k$, eigen volume $v_k$  and degeneracy $g_k$ is given by  $\phi_k(T) \equiv g_k ~ \phi(T,m_k) $  with 
\begin{equation}
\phi_k(T)   \equiv  \frac{g_k}{2\pi^2} \int\limits_0^{\infty}\hspace*{-0.0cm}p^2dp~
\exp{\textstyle \left[- \frac{(p^2~+~m_k^2)^{1/2}}{T} \right] } 
=  g_k \frac{m_k^2T}{2\pi^2}~{ K}_2 {\textstyle \left( \frac{m_k}{T} \right) }\, .
\end{equation}
\noindent
The mass-volume spectrum $\rho(m,v)$ is the generalization of the exponential mass spectrum 
introduced by Hagedorn \cite{Bugaev:Ref0}.  Similar to  the GBM, the QGBSTM1 and the QGBSTM2,  the 
FWM bags 
are assumed to have the hard core repulsion of the Van der Waals type which generates the suppression factor proportional to the  exponential $\exp(-sv)$  of  bag eigen volume $v$. 

The first term of Eq.~(\ref{Bugaev:Eq3}), $F_H$, represents the contribution of a finite number of low-lying
hadron states up to mass $M_0 \approx 2.5 $ GeV \cite{Bugaev:Ref2} which correspond to  different  flavors. 
This function has no $s$-singularities at
any temperature $T$ and can generate only  a simple pole of the isobaric partition, whereas  the mass-volume spectrum of the bags $F_Q(s,T)$ is   chosen to 
generate an essential  singularity $s_Q (T) \equiv p_Q(T)/T$ which defines  the QGP  pressure $p_Q(T)$  at zero baryonic densities 
\cite{Bugaev:Ref1}.  Very recently our group discovered  \cite{Bugaev:Ref5}   absolutely new way to generate the  rightmost singularity 
of the QGP which is a simple pole too. The latter is  achieved by matching  the deconfinement PT curve in the baryonic chemical potential and temperature plane with the curve of vanishing surface tension  coefficient  \cite{Bugaev:Ref5}.

Here I  use the parameterization  of  the  spectrum $\rho(m,v)$ introduced in  \cite{Bugaev:Ref2}. 
It assumes that 
\begin{equation}
\rho (m,v) =   \frac{ \rho_1 (v)  ~N_{\Gamma}}{\Gamma (v) ~m^{a+\frac{3}{2} } }
 \exp{ \textstyle \left[ \frac{m}{T_H}   -   \frac{(m- B v)^2}{2 \Gamma^2 (v)}  \right]  } \,, \quad {\rm with} \quad 
\rho_1 (v) = f (T)\, v^{-b}~ \exp{\textstyle \left[  -  \frac{\sigma(T)}{T} \, v^{\kappa}\right] }\,.
\label{Bugaev:Eq5} 
\end{equation}
In fact, the FWM  divides all hadrons into two groups depending on their width: the long (short) living   hadrons belong to $F_H$ (to $F_Q$).
As one can see from (\ref{Bugaev:Eq5}) the mass spectrum has a Hagedorn like parameterization and  the Gaussian attenuation  around the bag mass
$B v$ ($B$ is the mass density of a bag of a  vanishing width) with the volume dependent  Gaussian  width 
$\Gamma (v)$ or width hereafter. 
I  will distinguish it from the true width defined as 
$\Gamma_R = \alpha \, \Gamma (v)$ ($\alpha \equiv 2 \sqrt{2 \ln 2}\,$).
It is  necessary to stress    that the Breit-Wigner attenuation  of  a resonance mass cannot be used in the spectrum 
(\ref{Bugaev:Eq5}) because in case of finite width it would lead to a divergency of the mass integral in (\ref{Bugaev:Eq2}) 
or temperatures above  $T_H$ \cite{Bugaev:Ref2,Bugaev:Ref3, Bugaev:Ref4}. 

The normalization factor  is defined as
\begin{equation}\label{Bugaev:Eq6}
N_{\Gamma}^{-1}~ = ~ \int\limits_{M_0}^{\infty}
 \hspace*{-0.1cm} \frac{dm}{\Gamma(v)}
    \exp{\textstyle \left[  -   \frac{(m- B v)^2}{2 \Gamma^2 (v)}  \right] } \,.
\end{equation}
Such a  choice  of mass-volume spectrum  (\ref{Bugaev:Eq5}) is  a natural  extension   of early attempts 
 \cite{Bugaev:Ref10,Bugaev:Ref11,Bugaev:Ref12}
to explore the bag volume  as a statistically independent   degree of freedom to derive an internal  pressure 
of large bags. 
As it will be shown below   such a simple parameterization   not only allows  one  to resolve both of the conceptual problems discussed  above, but, what is very important,  it also  gives  us  an exactly solvable model. 

The volume spectrum in  (\ref{Bugaev:Eq5}) contains the surface free energy (${\kappa} = 2/3$) with the $T$-dependent 
surface tension which is parameterized as 
$\sigma(T) = \sigma_0 \cdot
\left[ \frac{ T_{c}   - T }{T_{c}} \right]^{2k + 1} $  ($k =0, 1, 2,...$) \cite{Bugaev:Ref1, Bugaev:Ref16},
where  $ \sigma_0 > 0 $ can be a smooth function of temperature.  For $T$ being not larger than   the (tri)critical temperature $T_{c}$ such a parameterization  is justified by the usual  cluster models 
like the FDM \cite{Bugaev:Ref13,Bugaev:Ref17} and SMM \cite{Bugaev:Ref18,Bugaev:Ref19,Bugaev:Ref20}, whereas 
the general consideration  for any  $T$   can be driven  by  the surface partitions of the Hills and Dales model 
\cite{Bugaev:Ref16}. In Refs.  \cite{Bugaev:Ref1,Bugaev:Ref5} it  is  argued  that at low baryonic densities 
the first order deconfinement phase transition degenerates into a cross-over just because of 
negative surface tension coefficient for $ T > T_{c} $. The other  consequences of  such a  
surface tension  parameterization and the discussion of the absence of the curvature free energy in 
(\ref{Bugaev:Eq5}) can be found in  \cite{Bugaev:Ref1, Bugaev:Ref4,Bugaev:Ref21}. 

The spectrum (\ref{Bugaev:Eq5}) has a simple form, but is rather general since both the width $\Gamma (v)$ and the bag mass density $B$ can be medium dependent. It clearly reflects the fact 
that the QGP bags are similar to   the ordinary  quasiparticles with the medium dependent characteristics (life-time, most probable values of  mass and volume). Now one can readily    derive the infinite bag pressure 
for two choices of the width: the volume independent width $\Gamma(v) \equiv \Gamma_0$ and 
the volume dependent width $\Gamma(v) \equiv \Gamma_1 = \gamma v^\frac{1}{2}$. 
As will be seen  below the latter resolves   both of the conceptual problems discussed earlier, whereas the former  parameterization  I use  for a comparison.

\section{Subthreshold Suppression of Large and Heavy QGP Bags }
Consider first the free bags. 
For large bag volumes ($v \gg M_0/B > 0$) the factor (\ref{Bugaev:Eq6})  can be
found as  $N_\Gamma \approx 1/\sqrt{2 \pi} $.   Similarly, one can show that  for heavy free bags  ($m \gg B V_0$, $V_0 \approx 1$ fm$^3$ \cite{Bugaev:Ref2},
ignoring the  hard core repulsion and thermostat)
\vspace*{-0.05cm}
\begin{equation}\label{Bugaev:Eq7}
 \rho(m)  ~ \equiv   \int\limits_{V_0}^{\infty}\hspace*{-0.1cm} dv\,\rho(m,v) ~\approx ~
\frac{  \rho_1 (\frac{m}{B}) }{B ~m^{a+\frac{3}{2} } }
\exp{ \textstyle \left[ \frac{m}{T_H}     \right]  } \,.
\end{equation}

\vspace*{-0.1cm}
\noindent
It originates in   the fact that  for heavy bags the 
Gaussian  in (\ref{Bugaev:Eq5}) acts like a Dirac $\delta$-function for
either choice of $\Gamma_0$ or $\Gamma_1$. 
Thus, the Hagedorn form of  (\ref{Bugaev:Eq7}) has a clear physical meaning and, hence, it  gives an additional argument in favor of the FWM. Also it gives an upper bound for the 
volume dependence of $\Gamma(v)$: the Hagedorn-like mass spectrum (\ref{Bugaev:Eq7}) can be derived, if for large $v$ the width  $\Gamma$ increases slower than $v^{(1 - \kappa/2)}= v^{2/3}$. 

Similarly to (\ref{Bugaev:Eq7}), one can estimate the width of heavy free bags  averaged over bag volumes and get  $ \overline{\Gamma(v) } \approx  \Gamma(m/B) $.
Thus, 
for  $\Gamma_1(v)$ the mass spectrum of heavy free QGP bags 
must be the Hagedorn-like one with the property that  heavy resonances have to develop   
the large  mean width $ \Gamma_1(m/B) = \gamma \sqrt{m/B}$ and, hence,  they 
are hardly   observable. 
Applying these arguments to the strangelets,
I conclude  that, if their mean volume is a few cubic fermis or larger, they  should survive  a  very short time,
which  is similar to the results of  Ref. \cite{Bugaev:Ref22}  predicting  an instability of such strangelets.

Note also that such a mean width is essentially different from both the linear mass dependence of string models  \cite{Bugaev:Ref23} and from an  exponential  form  of the nonlocal field theoretical models \cite{Bugaev:Ref24}.
Nevertheless, as it will be  demonstrated   while discussing the Regge trajectories,  
 the mean width $ \Gamma_1(m/B) $ leads to the linear 
Regge trajectory of  heavy free QGP bags for large values of the invariant mass squared.

Next let's   calculate  $F_Q(s,T)$ (\ref{Bugaev:Eq3}) for the  spectrum (\ref{Bugaev:Eq5}) performing the mass integration. There are, however, two distinct  possibilities, depending on the sign of the most probable mass: 
\begin{equation}\label{Bugaev:Eq8}
 \langle m \rangle ~ \equiv ~  B v + \Gamma^2 (v) \beta\,,\quad {\rm with} 
\quad \beta \equiv  T_H^{-1} - T^{-1} \,. 
\end{equation}
 %
 %
%\vspace*{-0.0cm}
\noindent
If {\boldmath 
$ \langle m \rangle > 0$} for $v \gg V_0$,  one can use the saddle point method
for mass integration to  find  the function~$F_Q (s,T)$
\begin{equation}\label{Bugaev:Eq9}
 F_Q^+ (s,T)   \approx \left[  \frac{T}{2\pi} \right]^{\frac{3}{2} }
\int\limits_{V_0}^{\infty}dv ~ \frac{ \rho_1(v) }{\langle m \rangle^a} ~\exp{\textstyle \left[  \frac{(p^+  - sT )v}{T}  \right]} \, 
\end{equation}
\noindent
and the pressure of large  bags 
\vspace*{-0.05cm}
\begin{equation}\label{Bugaev:Eq10}
p^+ \equiv T \left[ \beta B + \frac{\Gamma^2 (v)}{2 v} \beta^2 \right] \,, \quad {\rm for} \quad  \Gamma (v) = \Gamma_1 (v) \quad \Rightarrow  \quad  p^+ = T \left[ \beta B + \frac{1}{2}\gamma^2 \beta^2 \right] 
\end{equation}
%\noindent
To get  (\ref{Bugaev:Eq9}) one has to use  in (\ref{Bugaev:Eq3}) an asymptotic form  of the $K_2$-function  $\phi(T,m)\simeq (mT/2\pi)^{3/2}\exp(-m/T)$ for $m\gg~T$, 
{collect all terms with $m$ in the exponential, get a full square for $(m -  \langle m \rangle)$ 
and make the Gaussian integration.}

Since for $s  <  s_Q^*(T) \equiv p^+(v\rightarrow \infty)/T $ the integral  (\ref{Bugaev:Eq9}) diverges at  its upper limit, 
the  partition (\ref{Bugaev:Eq2}) has  an essential singularity that  corresponds to
the QGP pressure of  an  infinite  large   bag.  One concludes that
the width  $\Gamma$ cannot grow faster than $v^{1/2}$ for $v\rightarrow \infty$, otherwise $p^+(v\rightarrow \infty) \rightarrow \infty $ and  $F_Q^+ (s,T)$ diverges for any $s$.
Thus,  for {\boldmath $ \langle m \rangle > 0$} the phase structure of the FWM  with  $\Gamma (v) = \Gamma_1 (v)$ is similar to the QGBSTM1 \cite{Bugaev:Ref1}, but it can be also reformulated to have the phase structure of QGBSTM2  \cite{Bugaev:Ref5}
by proper choice of the surface tension temperature dependence and value of the index  $\tau > 2$.
Comparing the $v$ power of the exponential pre-factor in (\ref{Bugaev:Eq9}) to  the continuous 
volume spectrum of bags of the QGBSTM1 and  QGBSTM2,  one  finds  that 
$a + b \equiv \tau$.
The volume spectrum of bags  $F_Q^+ (s,T)$ (\ref{Bugaev:Eq9}) is of general nature  and  has a clear physical meaning.  Choosing different $T$ dependent functions $\langle m \rangle$ and $\Gamma^2 (v)$, one obtains  different equations of state.

It is  possible to use the spectrum (\ref{Bugaev:Eq9}) not only for infinite system volumes, but also for 
finite volumes $V \gg V_0$. In this case the upper limit of integration should be replaced by finite  $V$ 
(see Refs. \cite{Bugaev:Ref10, Bugaev:Ref11} for details).  This will change the singularities of partition 
 (\ref{Bugaev:Eq2}) to a set of simple poles  $ s_n^*(T)$ in the complex $s$-plane which are  defined by the same equation as for 
 $V \rightarrow \infty$.  Similarly to the finite $V$ solution of the GBM 
\cite{Bugaev:Ref10, Bugaev:Ref11},  it can be shown that for finite $T$ the FWM  simple poles may have  small positive or even negative real part which would lead to a non-negligible contribution of the QGP bags into the total spectrum  $F(s,T)$  (\ref{Bugaev:Eq3}).
In other words, if the spectrum (\ref{Bugaev:Eq9}), was the only volume spectrum of the QGP bags, then there would exist a finite (non-negligible) probability to find  heavy QGP bags ($m \gg M_0$)  in finite systems  at low temperatures  $T \ll T_c$.  
Therefore, using the results of   the finite volume GBM and SMM,  one should   conclude that the spectrum 
(\ref{Bugaev:Eq9}) itself
cannot   explain  the absence of  the QGP bags at  $T \ll T_c$  and, hence, an alternative explanation of this fact is required. 

Such an explanation corresponds to the case {\boldmath$ \langle m \rangle \le 0 $} for $v \gg V_0$.
From (\ref{Bugaev:Eq8}) one can see that  
for the volume dependent width $\Gamma (v) = \Gamma_1 (v) $ the most probable mass $ \langle m \rangle $ inevitably becomes negative at low $T$, if $0 < B < \infty$. 
 In this case the maximum of the Gaussian  mass distribution is located at  resonance masses 
$m = \langle m \rangle \le 0 $. This is true for any argument of the $K_2$-function
 in   $F_Q(s,T)$  (\ref{Bugaev:Eq3}). 
Since the lower limit of  mass integration  $M_0$ lies above $ \langle m \rangle $, then  only the tail  of  
the  Gaussian mass distribution  may contribute  into $F_Q(s,T)$. 
A  thorough  inspection of the integrand in $F_Q(s,T)$  shows 
that above $M_0$  it is strongly decreasing function of resonance mass and, hence, only the vicinity
of  the lower limit of  mass integration  $M_0$ sizably  contributes into $F_Q(s,T)$.
Applying 
the steepest descent method and  the $K_2$-asymptotic form   for $M_0 T^{-1} \gg 1$
one obtains
\begin{equation}\label{Bugaev:Eq11}
\hspace*{-0.3cm}F_Q^-(s,T) \hspace*{-0.05cm} \approx  \hspace*{-0.075cm} \left[  \frac{T}{2\pi} \right]^{\hspace*{-0.05cm}\frac{3}{2} } 
\hspace*{-0.15cm}
\int\limits_{V_0}^{\infty} \hspace*{-0.15cm} dv  \frac{ \rho_1(v) N_{\Gamma}\, \Gamma (v)\, \exp{\textstyle \left[  \frac{(p^-  - sT )v}{T}  \right]}
}{M_0^a\, [M_0 - \langle m \rangle  + a \, \Gamma^{2} (v)/ M_0 ]} \,, \quad {\rm with} \quad 
p^-\big|_{v \gg V_0}  = {\textstyle  \frac{T}{v} \left[  \beta M_0 -  \frac{(M_0 - Bv)^2}{2\, \Gamma^{2} (v)}  
 \right] }\end{equation}
\noindent
where $p^-$ defines  the pressure of  QGP bag.

It is necessary  to stress that  the last result requires $B>~0$ and it cannot be obtained for a weaker $v$-growth than  $\Gamma(v) = \Gamma_1(v)$. 
Indeed, if $B<0$, then the normalization factor (\ref{Bugaev:Eq6}) would not be $1/\sqrt{2 \pi}$, but would become 
$N_{\Gamma} \approx   [M_0 - \langle m \rangle]\, \Gamma^{-1} (v) \exp{\textstyle 
\left[    \frac{(M_0 - Bv)^2}{2\, \Gamma^{2} (v)}  \right]} $ and, thus,  it would cancel 
the leading  term in  pressure (\ref{Bugaev:Eq11}). Note, however, that  the  inequality 
{\boldmath$ \langle m \rangle \le  0 $} for all $v \gg V_0$ with positive $B$ and 
finite $p^-(v \rightarrow \infty)$ is possible
for  $\Gamma(v) = \Gamma_1(v)$ only. In this case the pressure of an infinite bag is
%
%\vspace*{-0.15cm}
\begin{equation}\label{Bugaev:Eq12}
p^-(v \rightarrow \infty)   =  {\textstyle  - T\frac{ B^2}{2\, \gamma^2 } }  \,.
\end{equation}

Also it is necessary to point out  that the only width $\Gamma(v) = \Gamma_1(v)$ 
does not lead to any divergency  of  the bag pressure in thermodynamic limit. 
This is clearly seen from Eqs. (\ref{Bugaev:Eq10}) and (\ref{Bugaev:Eq11}) since the multiplier 
$\Gamma^2(v)$ stands in the numerator  of the pressure (\ref{Bugaev:Eq10}), whereas 
in the pressure (\ref{Bugaev:Eq11}) it appears in the denominator.  Thus, if one chooses 
the different $v$-dependence  for the  width, then either $p^+$ or $p^-$  would 
diverge for the bag of  infinite  size.

The new outcome of this case with $B>0$ is that for $T < T_H/2$ \cite{Bugaev:Ref2,Bugaev:Ref3,Bugaev:Ref4} the spectrum 
(\ref{Bugaev:Eq11}) contains the lightest QGP bags having the  smallest volume since 
every term in the pressure $p^-$ (\ref{Bugaev:Eq11}) is negative.  The finite volume of the system is no longer 
important   because only  the  smallest bags survive in (\ref{Bugaev:Eq11}).
Moreover, if such bags are created, they would have mass about  $M_0$ and
the width about $\Gamma_1(V_0)$, and, hence, they would not be distinguishable 
from the usual low-mass hadrons. 
Thus, the case {\boldmath$ \langle m \rangle \le 0 $} with 
$B>0$ leads to the {\it subthreshold suppression of the QGP bags} at low temperatures,
since their most probable mass is below the mass threshold  $M_0$ of the spectrum $F_Q(s,T)$.   Note that such an effect cannot be derived within  any of  the GBM-kind models  proposed earlier.
The negative values of  $ \langle m \rangle $ that appeared in the 
expressions above 
serve as an indicator of a different  physical
situation comparing to $ \langle m \rangle > 0$, but    have no physical meaning since 
$ \langle m \rangle \le 0 $ does not enter   the main physical observable  $p^- $.

The obtained results give us a unique   opportunity to make a bridge between the
particle phenomenology, some experimental facts and the LQCD. 
For instance, if the most probable mass of the QGP bags is known along with the QGP pressure, one can estimate the width of these  bags directly from Eqs. (\ref{Bugaev:Eq10})
and  (\ref{Bugaev:Eq11}). 
To demonstrate the new possibilities 
let's now consider   several examples of the QGP  EOS  and relate them  to the above results. First, let's  study the possibility of  getting   the MIT bag model pressure 
 $p_{bag} \equiv \sigma T^4 - B_{bag} $  \cite{Bugaev:Ref6}  by  the  stable QGP bags, i.e. 
$\Gamma (v) \equiv 0$. Equating the pressures $p^+$ and $p_{bag}$, one finds that 
the Hagedorn temperature is  related to a bag constant 
$B_{bag} \equiv \sigma T^4_H$.  Then the mass density of such bags  $\frac{\langle m \rangle}{v} $ is identical to 
\begin{equation}\label{Bugaev:Eq13}
B  =  \sigma T_H (T  + T_H)(T^2 + T_H^2)\,,
\end{equation}
and, hence, it is
always positive. Thus, the MIT bag model  EOS  can be easily obtained within 
the FWM, but, as was discussed earlier, such bags should  be  observable.

The key point of the width estimation is based on the fact that  the low $T$ pressure (\ref{Bugaev:Eq12}) resembles  the linear  $T$ dependent  term in  the LQCD pressure reported  by several groups (for the full list of  references see 
\cite{Bugaev:Ref3}).  In \cite{Bugaev:Ref3} it is argued that  the ansatz 
\begin{equation}\label{Bugaev:Eq14}
B(T) =  \sigma T_H^2  (T^2 + TT_H + T_H^2) \,,
\end{equation}
not only gives the simplest possibility to fulfill all the necessary requirements to the mass density of QGP bags,
but it also  reproduces the LQCD pressure.  
Moreover, comparing ansatz (\ref{Bugaev:Eq14}) with the mass density (\ref{Bugaev:Eq13}) obtained  for the pure MIT bag model pressure, one can see  that they differ only  by a term  $\sigma T_H T^3$ 
which at  low   $T \le 0.5 \, T_H$  is a negligible  correction to (\ref{Bugaev:Eq14}). 
Therefore, 
for low $T$  the ansatz (\ref{Bugaev:Eq14})  looks  quite reasonable because in this region it corresponds to the mass density of the most popular EOS  of  modern  QCD phenomenology. 

Using  (\ref{Bugaev:Eq14}) it was possible to find out that 
the true width is independent for the number of quark  flavors  and of   the  color group number 
and is 
$\Gamma_R (V_0, T=0) \approx 1.22\,  V_0^{\frac{1}{2}} \, T_c^{\frac{5}{2}} \alpha \approx 600 $ MeV and
$\Gamma_R (V_0, T=T_H)   = \sqrt{12}\,\Gamma_R (V_0, T=0) \approx 2000$ MeV. 
These  estimates clearly demonstrate  that there is no way to detect the decays of such
short  living  QGP bags even,  if they are  allowed  by the subthreshold suppression.

\section{Asymptotic Behavior of the QGP Bag Regge Trajectories}

The behavior of the width of hadronic  resonances was extensively 
studied almost forty years ago  in the  Regge poles method
what was dictated by an intensive analysis of the strongly interaction dynamics in high energy hadronic 
collisions.
A lot of effort was put forward \cite{Bugaev:Ref25, Bugaev:Ref26, Bugaev:Ref28} to elucidate the asymptotical behavior of the resonance trajectories
$\alpha(S)$ for $|S| \rightarrow \infty$ ($S$ is an invariant mass square in the reaction).  Since the Regge trajectory  determines not only the mass of  resonances,
but their width as well, it is worth   to compare these results  with the  FWM predictions. 
Note that nowadays there is  great   interest in  the behavior of the  Regge  trajectories of higher resonances in  the  context  of the 
5-dimensional {string   theory holographically dual to QCD}  \cite{Bugaev:Ref29} which is known as  AdS/CFT.

In what follows I use the results of  
Ref. 
\cite{Bugaev:Ref28} which is based on the following most general assumptions:
(I) $\alpha(S)$ is an analytical function, having only the physical cut  from 
$S= S_0$ to $S = \infty$; (II)  $\alpha(S)$ is polynomially restricted at the whole physical sheet;  (III) there exists a finite limit of the phase trajectory at 
$S \rightarrow  \infty$.
Using these assumptions, it was possible to prove \cite{Bugaev:Ref28}  that  for
$S \rightarrow  \infty$ the upper bound  of  the Regge  trajectory asymptotics 
at the whole physical sheet 
is
\begin{equation}\label{Bugaev:Eq15}
\alpha_u (S) = - g^2_u \left[  - S   \right]^\nu \,,\quad {\rm with} \quad   \nu \le 1 \,,
\end{equation}
where the function  $g^2_u > 0$ should increase slower than any power 
in this limit  and   its  phase must  vanish at  $|S| \rightarrow ~\infty$.

On the other hand, in Ref. \cite{Bugaev:Ref28}  it was also shown that, if 
in addition to (I)-(III) 
one requires  that the   transition amplitude $T(s,t)$ is 
a  polynomially  restricted function of $S$ for all nonphysical $t > t_0 > 0$,
then the real part of the Regge tragectory  does not  increase  at 
$|S| \rightarrow ~\infty$ and  the trajectory behaves as 
\begin{equation}\label{Bugaev:Eq16}
\alpha_l (S) =  g^2_l \left[ - \left[  - S   \right]^{\frac{1}{2}} + C_l \right] \,,
\end{equation}
where  $g^2_l > 0$ and $C_l$ are some constants.  Moreover,  (\ref{Bugaev:Eq16}) defines the lower bound for the asymptotic behavior of the Regge trajectory \cite{Bugaev:Ref28}. The  expression (\ref{Bugaev:Eq16})   is a generalization of a well known  Khuri  result \cite{Bugaev:Ref30}. It means that for each family of hadronic  resonances  the Regge poles do not go beyond  some vertical line in the complex spin plane. 
In other words,  it means that in asymptotics $S \rightarrow + \infty$ the resonances become infinitely wide, i.e. they are moving out of the real axis of the proper angular momentum $J$ and, therefore,   there are only  a finite number of resonances in the corresponding transition amplitude.   At first glance  it seems that the huge deficit  of heavy hadronic resonances 
compared to the Hagedorn mass  spectrum  supports such a conclusion.
Since there is   a finite number of resonance families \cite{Bugaev:Ref31} it is impossible to  generate from them an exponential mass spectrum
and, hence, 
the Hagedorn mass spectrum   cannot exist 
for large resonance masses.  
Consequently, the GBM and its  followers  run  into  a deep trouble.  However, it was possible to show \cite{Bugaev:Ref3} that the FWM with   negative value of the most probable bag mass $ \langle m \rangle \le  0$ can help to resolve this problem as well. 

Note that the direct comparison of the FWM predictions with the 
Regge poles asymptotics is impossible because  the resonance mass and its 
width $\Gamma(v)$ are independent variables in the FWM. 
Nevertheless, one  can relate their  average values  and compare them to the results of 
Ref. \cite{Bugaev:Ref28}.  
{
To illustrate this statement, 
let's  recall the result  on the mean Gaussian  width of the free bags averaged with respect to their volume  
by the spectrum (\ref{Bugaev:Eq7})
(see  two paragraphs after Eq. (\ref{Bugaev:Eq7}) for details)  
\begin{equation}\label{Bugaev:Eq17}
\overline{\Gamma_1(v) } \approx  \Gamma_1(m/B) = \gamma
\sqrt{ \frac{m}{B} }\,. 
\end{equation}
Using the formalism of  \cite{Bugaev:Ref28}, it  can  be shown that 
at zero temperature
the free QGP bags of mass $m$ and mean resonance  width $\alpha\, \overline{\Gamma_1(v) } |_{T=0} \approx \alpha\,\gamma_0
\sqrt{ \frac{m}{B_0} }$  
precisely correspond  to  the following Regge trajectory 
\begin{equation}\label{Bugaev:Eq18}
 \hspace*{-0.1cm}
\alpha_r  (S) =  g^2_r [S + a_r (- S)^\frac{3}{4} ] \quad  {\rm with} \quad {a_r =const  < 0}\,,
\end{equation}
with $\gamma_0 \equiv \gamma (T=0)$ and  $B_0 \equiv B(T=0)$ from (\ref{Bugaev:Eq14}).
Indeed, using $S = |S| e^{i \, \phi_r}  $ in (\ref{Bugaev:Eq18}), and  expanding  the second term on the right hand side of   (\ref{Bugaev:Eq18}) and requiring ${\rm Im} \left[ \alpha_r (S) \right] = 0$, one finds the phase of  
physical trajectory (one of four roots of one fourth power in  (\ref{Bugaev:Eq18})), 
which is vanishing in the correct quadrant of the complex $S$-plane, and going to the complex energy plane $E = \sqrt{S} \equiv M_r - i \frac{\Gamma_r}{2}$, 
one can also  determine the mass  $M_r$ and the width $\Gamma_r$ 
\begin{equation}\label{Bugaev:Eq19}
{}  
\phi_r  (S) \rightarrow  \frac{a_r \sin \frac{3}{4}\pi  }{|S|^\frac{1}{4} }  \rightarrow 
0^-\,,  \quad 
M_r \approx |S|^\frac{1}{2} \, \quad  {\rm and } \quad   \Gamma_r \approx - |S|^\frac{1}{2}
\phi_r  (S) =  \frac{|a_r| |S|^\frac{1}{4}   }{  \sqrt{2} }  =  \frac{|a_r| M_r^\frac{1}{2}   }{  \sqrt{2} },
\end{equation}
 %
%%%\label{Bugaev:Eq19}
%
of a resonance belonging to the trajectory  (\ref{Bugaev:Eq18}).

Comparing the mass dependence of the width in (\ref{Bugaev:Eq19}) with the mean width of  free QGP bags 
(\ref{Bugaev:Eq17}) taken at $T=0$, it is natural  to identify them 
\begin{equation}\label{Bugaev:Eq20}
{}
a_r^{free} \approx - \alpha \,\gamma_0 \, \sqrt{\frac{2}{B_0}} = - 4  \, \gamma_0 \, \sqrt{\frac{\ln 2}{B_0}}  \,, 
\end{equation}  
and to deduce  that   the free QGP bags belong to the Regge trajectory (\ref{Bugaev:Eq18}).
Such a conclusion is in line both with the well established results on the linear  Regge trajectories of hadronic resonances \cite{Bugaev:Ref31} and
with  theoretical expectations  of  the dual resonance model \cite{Bugaev:Ref32}, the open string model 
\cite{Bugaev:Ref33,Bugaev:Ref23}, the closed string model \cite{Bugaev:Ref33} and  the   AdS/CFT \cite{Bugaev:Ref29}.
Such a property of the FWM also  gives   a very strong argument in favor of  both  the 
volume dependent width $\Gamma (v) = \Gamma_1(v)$ and  
the corresponding mass-volume spectrum of heavy bags 
(\ref{Bugaev:Eq5}).
}

Next I consider the second  way of averaging the mass-volume spectrum 
with respect to  the resonance  mass
\begin{equation}\label{Bugaev:Eq21}
{} \hspace*{-.8cm}
\overline{m} (v)  ~ \equiv ~    \int\limits_{M_0}^{\infty}\hspace*{-0.0cm} dm 
\int  \frac{d^3k}{(2\pi)^3}  \,\rho(m,v) ~ m ~ e^{- \frac{\sqrt{k^2 + m^2} }{T}} 
 \left[   \int\limits_{M_0}^{\infty}\hspace*{-0.cm} dm
\int  \frac{d^3k}{(2\pi)^3}  \,\rho(m,v) ~e^{- \frac{\sqrt{k^2 + m^2} }{T}}  \right]^{-1}
\, ,
\end{equation}
which is technically easier  than averaging with respect to the resonance volume. 
The latter  can be found in  \cite{Bugaev:Ref3}.

Using the results of Sect. 3  one can find the mean mass (\ref{Bugaev:Eq21})
for $T \ge  \,0.5\, T_H$ (or for $ \langle m \rangle \ge 0 $ ) to be equal to 
the most probable mass of  bag from which one determines the resonance width:
\begin{equation}\label{Bugaev:Eq22}
{}  \hspace*{-.25cm}
\overline{m} (v)   \approx  \langle m \rangle  \quad {\rm and} 
% 
%%%\label{mgammaII}
{}  \hspace*{.5cm}
\Gamma_R (v) \approx  2 \sqrt{2 \ln2 } ~ \Gamma_1 \hspace*{-.1cm} \left[ 
\frac{ \langle m \rangle  }{\textstyle B + \gamma^2 \beta}  \right] 
\hspace*{-.1cm}  = 2 \,\gamma \sqrt{ 
\frac{2 \ln 2 \,\langle m \rangle }{B + \gamma^2 \beta}  }  \,.
\end{equation}
These  equations lead to a vanishing  ratio 
$\frac{\Gamma_R}{\langle m \rangle} \sim  \langle m \rangle^{-\frac{1}{2}}$ in the limit $\langle m \rangle \rightarrow \infty$.
Comparing (\ref{Bugaev:Eq22})   with the mass and width  (\ref{Bugaev:Eq19}) of the resonances described by the 
Regge trajectory (\ref{Bugaev:Eq18}) and applying absolutely the same logic which was  used for the 
free QGP bags, I conclude 
that the location of the FWM heavy bags in the complex energy plane 
is identical to that one of resonances belonging to the trajectory 
(\ref{Bugaev:Eq18}) with 
\begin{equation}\label{Bugaev:Eq23}
{}  
\langle m \rangle \approx |S|^\frac{1}{2} \quad {\rm and } \quad 
a_r \approx  - 4 \gamma \, \sqrt{  \frac{\ln 2   }{  B + \gamma^2 \beta }  }
\,.
\end{equation}
The  most remarkable output of such a conclusion  is that   the medium dependent FWM  mass and  width of the extended QGP bags obey the upper  bound  for  the Regge
trajectory asymptotic  behavior   obtained for point-like hadrons 
\cite{Bugaev:Ref28}!

The extracted  values of the  resonance width coefficient along with the relation  (\ref{Bugaev:Eq14}) for  $B (T)$ allow us to estimate $a_r$  as 
\begin{equation}\label{Bugaev:Eq24}
a_r \approx  - 4  \, \sqrt{   \frac{2\,    T\, T_H \,  }{ 2\, T - T_H }   \ln 2 } \,. 
\end{equation}
This expression shows  that for $T \rightarrow T_H/2  + 0$  the asymptotic behavior 
(\ref{Bugaev:Eq18}) breaks down since the resonance width  diverges  at fixed $|S|$.  I think 
such a behavior can be studied at NICA (Dubna) and/or  FAIR (GSI, Darmstadt) energies. 
This can be seen from the following estimates.
From (\ref{Bugaev:Eq24}) it follows that  $a^2_r (T= T_H)  \approx  22.18 \, T_H $
and  $a^2_r (T\gg T_H)  \approx  11.09 \, T_H $. In other words,  for a typical value of 
the Hagedorn  temperature  $T_H \approx 190 $ MeV 
(\ref{Bugaev:Eq24}) gives a reasonable range of the invariant mass  $|S|^\frac{1}{2} \gg a^2_r (T= T_H)  \approx 4.21 $ GeV  and  $|S|^\frac{1}{2} \gg a^2_r (T \gg T_H)  \approx 2.1 $ GeV
for which  Eq.   (\ref{Bugaev:Eq18}) is true. 
But then at $|S|^\frac{1}{2} \approx  a^2_r (T= T_H)  \approx 4.21 $ GeV   the asymptotic behavior 
 (\ref{Bugaev:Eq18}) should be broken down and, hence, one may  see  the {\it  resonances widening} at fixed 
 $S$ values and decreasing $T$.

Now it is possible to find the spin of the FWM  resonances 
$
J  = {\rm Re} \, \alpha_r (\langle m \rangle^2) \, \, \approx   g_r^2  \,
 \langle m \rangle \left[ \langle m \rangle - \frac{a_r^2}{4} \right] \,,
$
which  has  a typical Regge behavior up to a small correction.
Such a property can also be  obtained   within  the dual resonance model \cite{Bugaev:Ref32},
within the models of  open  \cite{Bugaev:Ref33}  and  closed  string \cite{Bugaev:Ref33,Bugaev:Ref23}, and within 
 AdS/CFT \cite{Bugaev:Ref29}.
These models  support  the  relation between  the spin and mass  and justify  it.  Note, however, that 
in addition to the spin value  the FWM determines the width of hadronic  resonances. 
The latter  allows one  to predict   the  ratio 
of widths of two resonances having spins $J_2$ and $J_1$
and appearing at the same temperature $T$  to be  as follows
\begin{equation}\label{Bugaev:Eq25}
\Gamma_R \biggl[ \frac{\langle m \rangle\bigl|_{J_2}}{ (B + \gamma^2 \beta) } \biggl]
\left[ \Gamma_R \biggl[ \frac{\langle m \rangle\bigl|_{J_1}}{ (B + \gamma^2 \beta)} \biggl] \right]^{-1}
\approx 
\sqrt{v \bigl|_{J_2} }  
\left[  \sqrt{v \bigl|_{J_1}  }    \right]^{-1}
\approx \ \sqrt{\langle m \rangle\bigl|_{J_2} }  
\left[  \sqrt{\langle m \rangle\bigl|_{J_1}  }    \right]^{-1}
\approx   \left[  \frac{J_2}{J_1} \right]^\frac{1}{4}\,,
\end{equation}
which, perhaps, can  be tested at LHC.

Now let's  analyze   the low temperature regime, i.e. 
$T \le \,0.5 T_H$.  Using previously obtained results from  (\ref{Bugaev:Eq21}) one finds 
\begin{equation}\label{Bugaev:Eq26}
{} 
\overline{m} (v)  ~ \approx ~  M_0 \,,
\end{equation}
i.e.
the mean mass is volume independent. Taking the limit $v \rightarrow \infty$, one  gets  the ratio $\frac{\Gamma(v)}{\overline{m} (v)} \rightarrow  \infty $ which closely resembles  the case of  the  lower bound  of the Regge trajectory asymptotics  (\ref{Bugaev:Eq16}). 
Similarly to the analysis of  high temperature regime, from (\ref{Bugaev:Eq16}) one can find  the trajectory  phase and then  the resonance  mass $M_r$ and 
its width $\Gamma_r$ 
\begin{equation}\label{Bugaev:Eq27}
{}  
\phi_r  (S) \rightarrow - \pi + \frac{2 |C_l |  |\sin (\arg C_l) |  }{|S|^\frac{1}{2} }  \,, \quad 
M_r \approx   |C_l |  |\sin (\arg C_l) | \quad {\rm and} \quad
\Gamma_r \approx  2 |S|^\frac{1}{2} \,. 
\end{equation}
Again comparing the  averaged  masses  and width of FWM resonances 
with  their counterparts in  (\ref{Bugaev:Eq27}), one  finds a   similar behavior in the 
limit of large  width of resonances.  Therefore, I 
conclude   that 
at low temperatures the FWM obeys the lower bound of the Regge trajectory asymptotics of  \cite{Bugaev:Ref28}.

The above  estimates  demonstrate  that at any temperature  the FWM QGP bags can be regarded as the medium induced Reggeons which at $T \le \,0.5 T_H$ 
(i.e. for  $\langle m \rangle \le 0$) belong to the Regge trajectory   (\ref{Bugaev:Eq16}) and otherwise they are described by the  trajectory (\ref{Bugaev:Eq18}).
Of course, both of the trajectories (\ref{Bugaev:Eq16}) and  (\ref{Bugaev:Eq18})
are valid in the asymptotic $|S| \rightarrow \infty$, but the most remarkable  fact is that,
to my  knowledge, the FWM gives us the first example of a model which reproduces both  of these trajectories and, thus, obeys both bounds of the Regge asymptotics.  Moreover, since the FWM contains the Hagedorn-like mass spectrum at any temperature, the subthreshold suppression of QGP bags removes the contradiction between the Hagedorn  ideas on the exponential mass spectrum of hadrons   and the Regge
poles method in the low temperature domain! Furthermore, the FWM opens a possibility to apply the Regge poles  method to a variety of processes in a strongly interacting  matter  and account, at least partly,  for 
some of the medium effects.

\section{Conclusions}  

Here I present  the novel statistical approach to study  the QGP bags with  medium dependent width. 
It is a further extension of the ideas formulated in \cite{Bugaev:Ref34}.
I  argue   that the volume dependent width of the QGP bags $\Gamma (v) = \gamma\, v^\frac{1}{2}$ leads to the Hagedorn mass spectrum of heavy bags.  Such  behavior of a width  allows us to explain a huge  deficit of heavy hadronic resonances in the experimental mass spectrum compared to the Hadegorn model 
predictions.  The key point of our treatment   is the presence of Gaussian  attenuation of bag mass. 

Then it is shown that  the Gaussian mass attenuation also  allows one  to ``hide''  the heavy  QGP bags for 
$T \le  \, 0.5 \,   T_H$  by their {\it subthreshold suppression}. The latter occurs due to the fact 
that at low temperatures the most probable mass of heavy bags $\langle m \rangle$  becomes negative and, hence, is below the lower cut-off  $M_0$ of
the continuous mass spectrum. 
Consequently,  only the lightest  bags of mass about $M_0$ and of  smallest volume $V_0$
may contribute into the resulting spectrum, but such QGP bags will be indistinguishable  
from the low-lying  hadronic resonances with the short life-time. 
On the other hand the large minimal width, about 600 MeV,  of  bags being  heavier than $M_0$ and large than 
$V_0$ prevents their experimental behavior for $T  >  \, 0.5 \,   T_H$, even, if  they are allowed by the 
subthreshold suppression. 
Thus, the FWM naturally   explains  the absence of directly observable QGP bags and strangelets  in the high energy nuclear 
and elementary particle collisions even as metastable states in hadronic phase.

Using the formalism of \cite{Bugaev:Ref28}  it was   shown  that  the average mass and width of  heavy/large   free QGP bags  belong to the linear Regge trajectory (\ref{Bugaev:Eq18}).
Similarly, it was found that
at  hight temperatures the average mass and width of the QGP bags  behave in accordance with  the upper bound of the Regge trajectory asymptotics (\ref{Bugaev:Eq15}) (linear trajectory), whereas at low temperatures they obey  the lower bound 
of   the Regge trajectory asymptotics (\ref{Bugaev:Eq16}) (square root trajectory). 
Such  results  create  a new look onto the large and/or heavy   QGP bags as the medium induced  Reggeons
and provide us with   an  alternative to the AdS CFT  picture on the QGP bags.  
I would like to stress that it can be used not only to describe the deconfinement or cross-over and the corresponding phases, but for the metastable states of strongly interacting matter as well. 
Also as shown above such a
coherent  picture 
not only introduces  new time scale into the heavy ion physics, but  also naturally explains the existence 
of  the tricritical  or  critical QCD endpoint  due to the vanishing of  the surface tension coefficient. 
Therefore, I am sure  that further development of such a rich  direction will lead to the major shift 
of the low energy paradigm of heavy ion physics and will shed light on the modification of the QCD (tri)critical endpoint 
properties in  finite systems. 

%%%KKK

\acknowledge{Acknowledgements}{
The research made in this work  
was supported in part   by the Program ``Fundamental Properties of Physical Systems 
under Extreme Conditions''  of the Bureau of the Section of Physics and Astronomy  of
the National Academy of Science of Ukraine.
This work was also supported in part by the Fundamental Research State Fund of Ukraine,
Agreement No F28/335-2009 for the Bilateral project FRSF (Ukraine) -- RFBR (Russia).
}

%Do not modify the rest of the file

\end{article}

\label{pgs1}
\end{document}